\documentclass[prd,preprintnumbers,superscriptaddress,twocolumn,epsf,nofootinbib,showpacs]{revtex4} 
 
\usepackage{amsfonts,amssymb,amsmath} 
\usepackage{mathrsfs} 
\usepackage{color} 
\usepackage{graphicx}
\usepackage{dcolumn}
\usepackage{bm}

\begin{document} 

\title{CMB from a Gauss-Bonnet-induced de Sitter fixed point} 
\author{Shinsuke Kawai} 
\email{kawai@skku.edu} 
\affiliation{
	Department of Physics, 
	Sungkyunkwan University, 
	Suwon 16419, Republic of Korea}
\author{Jinsu Kim}
\email{jinsu.kim@cern.ch}
\affiliation{
	Theoretical Physics Department,
	CERN,
	1211 Geneva 23, Switzerland} 
\date{\today} 
\preprint{CERN-TH-2021-075}

\begin{abstract} 
In the gravitational effective theories including higher curvature terms, cosmological solutions can have nontrivial de Sitter fixed points.
We study phenomenological implications of such points, focusing on a theory in which a massive scalar field is nonminimally coupled to the Euler density.
We first analyze the phase portrait of the dynamical system and show that the fixed point can be a sink or a saddle, depending on the strength of the coupling.
Then, we compute the perturbation spectra generated in the vicinity of the fixed point in order to investigate whether the fixed point may be considered as cosmic inflation.
We find parameter regions that are consistent with the cosmological data, given that the anisotropies in the cosmic microwave background are seeded by the fluctuations generated near the fixed point.
Future observation may be used to further constrain the coupling function of this model.
We also comment briefly on the swampland conjecture.
\end{abstract}
 
\maketitle

\section{Introduction}

Realizing a de Sitter-like solution in a consistent theory of high energy physics is known to be difficult.
Slow-roll inflation is not possible in supergravity with a canonical K\"ahler potential, a generic superpotential, and F-term supersymmetry breaking.
This is known as the supergravity $\eta$ problem \cite{Dine:1995uk,Dine:1995kz,Copeland:1994vg,Lyth:2004nx} (see also reviews \cite{Yamaguchi:2011kg,Mazumdar:2010sa}).
In string theory, there exists a powerful no-go theorem applicable to generic compactification scenarios \cite{Maldacena:2000mw}. 
See, e.g., Refs. \cite{McAllister:2007bg,Baumann:2014nda} for reviews.
Recently, broadly in quantum gravity, there have been a lot of activities under the name of the swampland program 
\cite{Vafa:2005ui,ArkaniHamed:2006dz} (see, e.g., Ref.~\cite{vanBeest:2021lhn} for a recent review).
One of the swampland conjectures precludes simple realization of potential-driven slow-roll type inflation.
These theorems and conjectures are important for phenomenological model building, since UV completeness is a compelling guiding principle.
The difficulty of realizing the traditional type of inflating spacetime motivates us to seek for alternative de Sitter-like solutions in a broader context.
The Einstein gravity is by no means a complete theory of gravity as higher curvature terms can arise from the renormalization of stress-energy tensors as well as in string perturbation theory.
Such terms are not included for example in the postulates of the no-go theorem \cite{Maldacena:2000mw}.
In this paper, we will be interested in the type of solutions that are realized by the balance of a scalar potential term and a higher curvature term.

In order to avoid difficulties such as the Ostrogradski instability and to make our discussions simple, we focus on a well-behaved gravity theory in which a scalar field $\varphi$ is coupled to the four-dimensional Euler density (also called the Gauss-Bonnet term in the literature)
$R_{\rm GB}^2\equiv R^2-4R_{\mu\nu}R^{\mu\nu}+R_{\mu\nu\rho\sigma}R^{\mu\nu\rho\sigma}$.
This type of correction to the Einstein gravity is expected for example in inflationary effective field theory \cite{Weinberg:2008hq}.
We consider the action, 
{\small
\begin{align}\label{eqn:action} 
	S=\!\!\int d^4 x\sqrt{-g}\Big\{\frac{M_{\rm P}^2}{2}R 
	-\frac{1}{2}(\partial\varphi)^2-V(\varphi)
	-\frac{\xi(\varphi)}{16}R_{\rm GB}^2 
	\Big\} \,,
\end{align}
}where $M_{\rm P}\equiv 1/\sqrt{8\pi G}=2.44\times 10^{18}$ GeV is the reduced Planck mass, $V(\varphi)$ is the scalar potential, and $\xi(\varphi)$ is the coupling function.
When the scalar field $\varphi$ is stabilized at $V(\varphi)=0$, Eq.~\eqref{eqn:action} becomes the Einstein-Hilbert action as the last term is topological for constant $\xi(\varphi)$.
Based on the action \eqref{eqn:action}, various cosmological models have been studied by many authors, with different assumptions on the potential $V(\varphi)$ and the coupling function $\xi(\varphi)$. 
This type of models appeared in the study of cosmological solutions that are free from the initial singularity \cite{Antoniadis:1993jc,Rizos:1993rt}.
Phenomenological effects, for example, on the lepton asymmetry generated by polarized gravitational waves have been discussed in Ref. \cite{Kawai:2017kqt}.
The Gauss-Bonnet term has also been studied as a source of dark energy \cite{Nojiri:2005vv,Koivisto:2006xf,Nojiri:2007te,Amendola:2007ni,Granda:2014zea}.
It has been pointed out \cite{Satoh:2007gn,Satoh:2008ck,Guo:2009uk,Guo:2010jr,Satoh:2010ep,Yi:2018gse,Nojiri:2019dwl,Odintsov:2019clh,Odintsov:2020sqy,Pozdeeva:2020shl,Oikonomou:2020sij,Pozdeeva:2020apf,Oikonomou:2020oil,Odintsov:2020xji,Odintsov:2020mkz,Oikonomou:2020tct} that the Gauss-Bonnet term, when included as a correction term to the traditional slow roll inflation models, may change the predicted spectrum of the primordial fluctuations. 
In particular, inflationary models that are disfavored by the latest Planck data \cite{Akrami:2018odb} may become compatible with the observation when the Gauss-Bonnet term with some suitably chosen coupling function $\xi(\varphi)$ is included.
For example, Ref. \cite{Satoh:2008ck} analyzes inflation with the quadratic $V \propto \varphi^2$ and the quartic $V \propto \varphi^4$ scalar potential.
These models are strongly disfavored by the recent observational data, but they can be made compatible if a Gauss-Bonnet term with the exponential coupling is included.
Similar conclusions were reached in other studies.

In this paper, our interest is in the fixed points of the cosmological solutions and possible phenomenological interpretation of them.
Unlike other studies, we leave the coupling function $\xi(\varphi)$ arbitrary, although we make simplifying assumptions on the potential $V(\varphi)$ in order to make our analysis concrete.
Ideally, the potential $V(\varphi)$ and the coupling $\xi(\varphi)$ are expected to be determined by some underlying theoretical model, such as string theory with a particular compactification scenario.
Our approach in this paper, in contrast, is phenomenological;
we consider the coupling function $\xi(\varphi)$ to be constrained by cosmological data.

In the next section, we analyze the cosmological model as a dynamical system.
Possible fixed points are classified, and their properties are examined.
Among the fixed points, de Sitter-like points are of phenomenological interest.
In Sec.~\ref{sec:USR}, we study the behavior of cosmological solutions near a de Sitter-like fixed point. 
Cosmological perturbation theory near the de Sitter-like fixed point is formulated in Sec.~\ref{sec:pert}.
Constraints on the model parameters by cosmological data are discussed in Sec.~\ref{sec:constraints}.
We conclude this paper with comments in Sec.~\ref{sec:concl}.
Two appendixes collect technicalities.

\section{Dynamical system}\label{sec:dynsys} 
For a spatially flat Friedmann-Lema\^{i}tre-Robertson-Walker spacetime, the Friedmann equation and the equation of motion for the scalar field are given by
{\small
\begin{gather}
3M_{{\rm P}}^2 H^2 = \frac{\dot{\varphi}^2}{2} + V + \frac{3}{2}H^3\xi_{,\varphi}\dot{\varphi}
\,,\\
\ddot{\varphi} + 3H\dot{\varphi} + V_{,\varphi} + \frac{3}{2}H^2\left(\dot{H}+H^2\right)\xi_{,\varphi} = 0
\,,
\end{gather}
}where $H$ is the Hubble parameter, the dot denotes the cosmic time derivative $\dot{}\equiv d/dt$, and ${}_{,\varphi} \equiv d/d\varphi$.
For a fixed point, $\dot{\varphi}=\ddot{\varphi}=\dot{H}=0$, and the above equations become
{\small
\begin{gather}
3M_{{\rm P}}^2H^2 = V
\,,\label{eqn:FPeqn1}\\
V_{,\varphi}+\frac{3}{2}H^4\xi_{,\varphi} = 0\label{eqn:FPeqn2}
\,.
\end{gather}
}The first equation indicates that $V$ is nonnegative, and thus, the fixed point is either flat Minkowski ($V=0$) or pure de Sitter ($V>0$).
In the absence of the Gauss-Bonnet coupling, $\xi=0$, the pure de Sitter fixed point is realized by a positive extremum of the scalar potential. When there is a Gauss-Bonnet coupling, there exists a nontrivial fixed point where the potential is not flat, provided that $V_{,\varphi}$ and $\xi_{,\varphi}$ have opposite signs. 

The Friedmann equation is cubic in the Hubble parameter $H$. 
We are interested in the solution that is real, positive and nonsingular in the limit $\xi\to 0$.
It is convenient to use Vieta's solution,
{\small
\begin{align}
H = \frac{\sqrt{\dot{\varphi}^2/2+V}}{2M_{{\rm P}}}
\sec\left(
\frac{1}{3}\arccos\left(
-\frac{3}{4}\frac{\xi_{,\varphi}\dot{\varphi}}{M_{{\rm P}}}
\sqrt{\frac{\dot{\varphi}^2/2+V}{M_{{\rm P}}^4}}
\right)
\right)\,.
\end{align}
}From the equations of motion, we have
{\small
\begin{align}
\frac{d}{dt}\varphi &=
\dot{\varphi}\,,\\
\frac{d}{dt}\dot{\varphi} &=
\frac{1}
{4H\xi_{,\varphi}\dot{\varphi}-3H^4\xi_{,\varphi}^2-8M_{{\rm P}}^2}
\bigg[
8M_{{\rm P}}^2V_{,\varphi}
+12H^4M_{{\rm P}}^2\xi_{,\varphi}
\nonumber\\
&\qquad
-\left(
18H^2\xi_{,\varphi}-3H^4\xi_{,\varphi}\xi_{,\varphi\varphi}
\right)\dot{\varphi}^2
\nonumber\\
&\qquad
-\left(
9H^5\xi_{,\varphi}^2-24M_{{\rm P}}^2H+4H\xi_{,\varphi}V_{,\varphi}
\right)\dot{\varphi}
\bigg]
\,,\\
\frac{d}{dt}H &=
\frac{\left(
2H^2\xi_{,\varphi\varphi}-4
\right)\dot{\varphi}^2
-8H^3\xi_{,\varphi}\dot{\varphi}-2H^2\xi_{,\varphi}V_{,\varphi}-3H^6\xi_{,\varphi}^2}
{3H^4\xi_{,\varphi}^2-4H\xi_{,\varphi}\dot{\varphi}+8M_{{\rm P}}^2}
\,,
\end{align}
}which define a dynamical system for three variables $(\varphi,\dot{\varphi},H)$. Due to the Hamiltonian constraint, the phase space is two dimensional.

To our knowledge, the first dynamical-system analysis of cosmological solutions in Gauss-Bonnet gravity appeared in Ref.~\cite{Koivisto:2006ai} (see also Refs.~\cite{Neupane:2006dp,Kim:2013xu}), and there has been recent revival of interest, e.g., Refs.~\cite{Granda:2017oku,Oikonomou:2017ppp,Pozdeeva:2019agu,Chatzarakis:2019fbn,Vernov:2021hxo}.
A recent review on this topic is Ref.~\cite{Bahamonde:2017ize}.
In the following, for concreteness, we focus on the quadratic inflaton potential,
{\small
\begin{align}
V(\varphi) = \frac{1}{2}m^2\varphi^2\,.
\end{align}
}The dynamical-system analysis with the quadratic potential is also performed in Refs. \cite{Granda:2017oku,Pozdeeva:2019agu}, where unstable de Sitter solutions are found. In Ref. \cite{Pozdeeva:2019agu}, for example, the authors used the effective potential approach rather than the fixed point approach, which we shall pursue. Furthermore, specific forms for the coupling function $\xi$ are considered. In Ref. \cite{Granda:2017oku}, fixed points were analyzed with specific forms for the coupling function with the focus on the late-time Universe. We, on the other hand, are interested in the early-time Universe with a generic form of the coupling function $\xi$.

It is convenient to work with dimensionless variables. We introduce
{\small
\begin{align}\label{eqn:DimlessVars}
\bar{t} \equiv mt \,,\;\;
\bar{H} \equiv \frac{H}{m} \,,\;\;
\bar{\varphi} \equiv \frac{\varphi}{M_{{\rm P}}} \,,\;\;
\mathring{\bar{\varphi}} \equiv \frac{d\bar{\varphi}}{d\bar{t}}\,,\;\;
\bar{\xi} \equiv \frac{m^2}{M_{{\rm P}}^2}\xi\,.
\end{align}
}In terms of the dimensionless variables, the Hubble parameter reads
{\small
\begin{align}
\bar{H} = \frac{\sqrt{\mathring{\bar{\varphi}}^2+\bar{\varphi}^2}}{2\sqrt{2}}
\sec\left(
\frac{1}{3}\arccos\left(
-\frac{3}{4\sqrt{2}}\bar{\xi}_{,\bar{\varphi}}\mathring{\bar{\varphi}}
\sqrt{\mathring{\bar{\varphi}}^2+\bar{\varphi}^2}
\right)
\right)\,,
\label{eqn:Hbark}
\end{align}
}while the equations for the dynamical system are given by
{\small
\begin{align}
\frac{d}{d\bar{t}}\bar{\varphi} &=
\mathring{\bar{\varphi}}\,,\\
\frac{d}{d\bar{t}}\mathring{\bar{\varphi}} &=
\frac{1}
{4\bar{H}\bar{\xi}_{,\bar{\varphi}}\mathring{\bar{\varphi}}-3\bar{H}^4\bar{\xi}_{,\bar{\varphi}}^2-8}
\bigg[
8\bar{\varphi}
+12\bar{H}^4\bar{\xi}_{,\bar{\varphi}}
\nonumber\\
&\quad
-\left(
18\bar{H}^2\bar{\xi}_{,\bar{\varphi}}-3\bar{H}^4\bar{\xi}_{,\bar{\varphi}}\bar{\xi}_{,\bar{\varphi}\bar{\varphi}}
\right)\mathring{\bar{\varphi}}^2
\nonumber\\
&\quad
-\left(
9\bar{H}^5\bar{\xi}_{,\bar{\varphi}}^2-24\bar{H}+4\bar{H}\bar{\xi}_{,\bar{\varphi}}\bar{\varphi}
\right)\mathring{\bar{\varphi}}
\bigg]
\,,\\
\frac{d}{d\bar{t}}\bar{H} &=
\frac{[2\bar{H}^2\bar{\xi}_{,\bar{\varphi}\bar{\varphi}}-4]\mathring{\bar{\varphi}}^2
-8\bar{H}^3\bar{\xi}_{,\bar{\varphi}}\mathring{\bar{\varphi}}-2\bar{H}^2\bar{\xi}_{,\bar{\varphi}}\bar{\varphi}-3\bar{H}^6\bar{\xi}_{,\bar{\varphi}}^2}
{3\bar{H}^4\bar{\xi}_{,\bar{\varphi}}^2-4\bar{H}\bar{\xi}_{,\bar{\varphi}}\mathring{\bar{\varphi}}+8}
\,.
\end{align}
}

At a fixed point, $\bar{\varphi} = \bar{\varphi}_*$, Eqs.~\eqref{eqn:FPeqn1} and \eqref{eqn:FPeqn2} give
{\small
\begin{align}
\bar{\varphi}_*
\left(
24+\bar{\varphi}_*^3\bar{\xi}_{,\bar{\varphi}_*}
\right) = 0\,.
\end{align}
}The trivial fixed point is $\bar{\varphi}_* = 0$, and the nontrivial fixed point satisfies
{\small
\begin{align}\label{eqn:nontrivialFP}
24+\bar{\varphi}_*^3\bar{\xi}_{,\bar{\varphi}_*} = 0\,.
\end{align}
}The characteristics of the fixed points are captured by the linear stability study.

\subsection{Linear stability analysis}

To study the dynamics near a fixed point, we linearize the equations about the fixed point.
We first expand the Gauss-Bonnet coupling function around the fixed point $\bar{\varphi}_*$,
{\small
\begin{align}
\bar{\xi}(\bar{\varphi}) = \bar{\xi}_0 + \bar{\xi}_1(\bar{\varphi} - \bar{\varphi}_*)
+\frac{1}{2}\bar{\xi}_2(\bar{\varphi} - \bar{\varphi}_*)^2
+\cdots
\end{align}
}Note that the nontrivial fixed point is given from Eq.~\eqref{eqn:nontrivialFP} as
{\small
\begin{align}\label{eqn:FPxi1}
\bar{\varphi}_* = \left(-\frac{24}{\bar{\xi}_1}\right)^{1/3}\,.
\end{align}
}

\subsubsection{Trivial fixed point}
The trivial fixed point is at $(\bar{\varphi},\mathring{\bar{\varphi}}) = (0,0)$. We redefine variables near the trivial fixed point,
{\small
\begin{align}
\bar{\varphi} = 0 + x
\,,\qquad
\mathring{\bar{\varphi}} = 0 + y\,,
\end{align}
}with $x, y \sim \mathcal{O}(\epsilon)$ and $\epsilon \ll 1$. To the lowest order in $\epsilon$, from Eq.~\eqref{eqn:Hbark}, we find
{\small
\begin{align}
\bar{H} = \frac{\sqrt{x^2+y^2}}{\sqrt{6}} + \mathcal{O}(\epsilon^3)\,.
\end{align}
}Using this, we see that the dynamical system becomes
{\small
\begin{align}
\frac{dx}{d\bar t} &= y \,,\\
\frac{dy}{d\bar t} &= -x -3\bar{H}_* y +\mathcal{O}(\epsilon^4)\,,
\end{align}
}or, in the matrix form,
{\small
\begin{align}
\frac{d}{d\bar t}
\left(\begin{array}{c}
{x} \\ {y}
\end{array}\right) = \left(\begin{array}{cc}
0 & 1 \\
-1 & -3\bar{H}_*
\end{array}\right)\left(\begin{array}{c}
x \\ y
\end{array}\right)\,,
\end{align}
}where $\bar{H}_* \equiv \sqrt{x^2+y^2}/\sqrt{6}$. The eigenvalues of the linearization matrix are given by
{\small
\begin{align}
-\frac{3}{2}\bar{H}_* \pm i\,.
\end{align}
}These are complex with negative real parts, representing a spiral sink, which is expected for the Minkowski vacuum in inflationary cosmology.

\subsubsection{Nontrivial fixed point}
\begin{figure*}[ht!]
\centering
\includegraphics[scale=0.45]{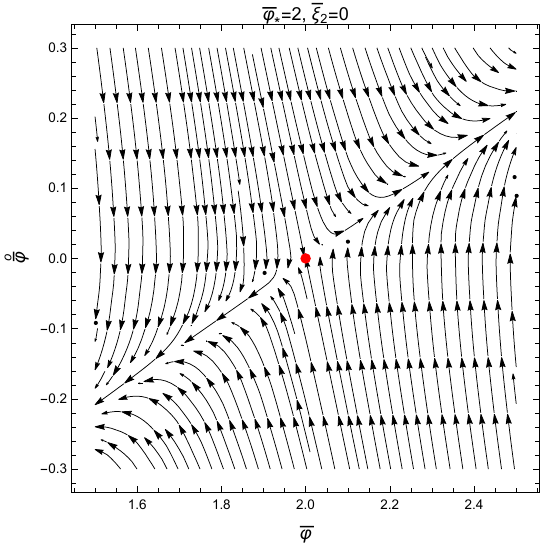}
\includegraphics[scale=0.45]{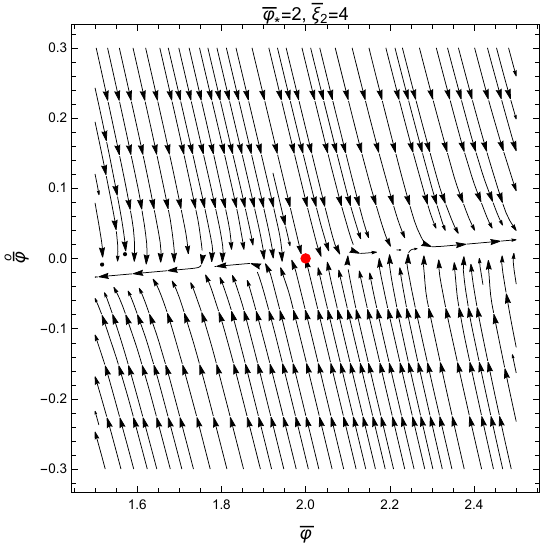}
\includegraphics[scale=0.45]{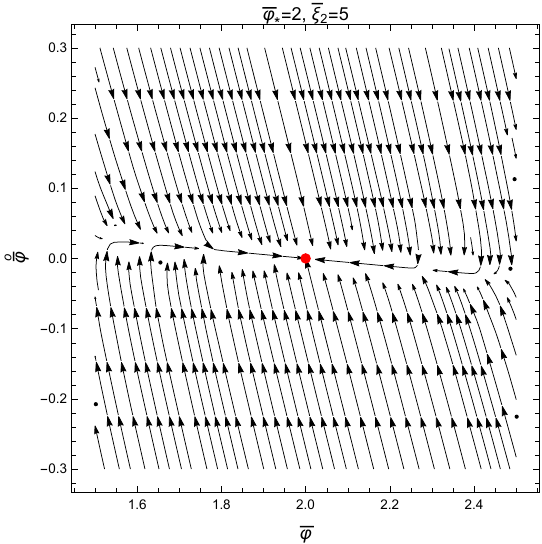}
\includegraphics[scale=0.45]{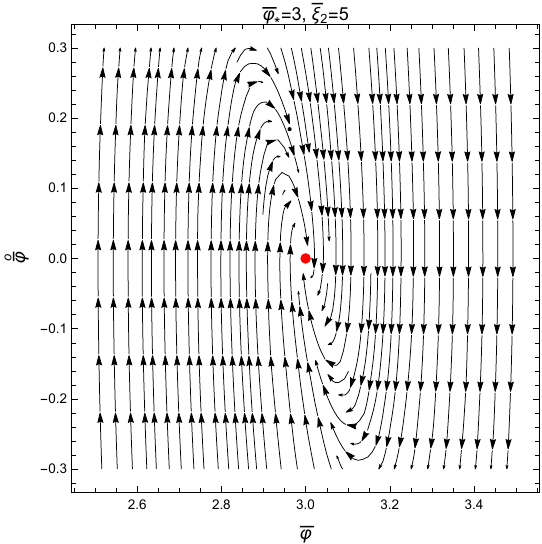}
\caption{Phase portraits for $\{\bar{\varphi}_*,\bar{\xi}_2\}=\{2,0\}$, $\{2,4\}$, $\{2,5\}$, and $\{3,5\}$ from left to right, respectively, showing that the nontrivial fixed points (red points) are a saddle point, a saddle point, a stable point, and a stable spiral point.}
\label{fig:phaseportraits}
\end{figure*}

Near the nontrivial fixed point $(\bar{\varphi},\mathring{\bar{\varphi}}) = (\bar{\varphi}_*,0)$, we may introduce variables $x$ and $y$ by
{\small
\begin{align}
\bar{\varphi} = \bar{\varphi}_* + x
\,,\qquad
\mathring{\bar{\varphi}} = 0 + y\,.
\label{eqn:xANDy}
\end{align}
}Up to the first order in $\epsilon$, the dimensionless Hubble parameter is
{\small
\begin{align}
\bar{H} \approx \frac{|\bar{\varphi}_*|}{\sqrt{6}}
+\frac{{\rm sgn}(\bar{\varphi}_*)}{\sqrt{6}}x
-\frac{1}{\bar{\varphi}_*}y
\,,
\label{eqn:Hxy}
\end{align}
}and the linearized equations for the dynamical system are given by
{\small
\begin{align}
\frac{dx}{d\bar t} &= y \,,\label{eqn:eomxy1}
\\
\frac{dy}{d\bar t} &=
\frac{\bar{\varphi}_*^2}{6+\bar{\varphi}_*^2}\left(
3 - \frac{\bar{\xi}_2 \bar{\varphi}_*^4}{24}
\right)x - \sqrt{\frac{3}{2}}|\bar{\varphi}_*| y
\,,
\label{eqn:eomxy2}
\end{align}
}where we used that $\bar{\xi}_1 = -24/\bar{\varphi}_*^3$.
In the matrix form,
{\small
\begin{align}
\frac{d}{d\bar t}
\left(\begin{array}{c}
{x} \\ {y}
\end{array}\right) = \left(\begin{array}{cc}
0 & 1 \\
\frac{\bar{\varphi}_*^2}{6+\bar{\varphi}_*^2}\left(
3 - \frac{\bar{\xi}_2 \bar{\varphi}_*^4}{24}
\right) & - \sqrt{\frac{3}{2}}|\bar{\varphi}_*|
\end{array}\right)\left(\begin{array}{c}
x \\ y
\end{array}\right)\,.
\end{align}
}The eigenvalues of the linearization matrix are
{\small
\begin{align}
\lambda_{\pm} = \frac{\sqrt{6}|\bar{\varphi}_*|}{12}\left(
-3 \pm \sqrt{\frac{126+9\bar{\varphi}_*^2-\bar{\xi}_2 \bar{\varphi}_*^4}{6+\bar{\varphi}_*^2}}
\right)\,.
\end{align}
}We may then classify the behavior of the fixed point into the following four cases:

\begin{enumerate}
  \item ${\bar{\xi}_2 \le 0}$\\
  In this case, both of $\lambda_{\pm}$ are real and $\lambda_{-} < 0 < \lambda_{+}$. Therefore, the nontrivial fixed point is a saddle point.
  \item ${\bar{\xi}_2 > 0,\quad |\bar{\varphi}_*|< \left(\frac{72}{\bar{\xi}_2}\right)^{1/4}}$\\
  We still have real $\lambda_{\pm}$ with  $\lambda_{-} < 0 < \lambda_{+}$.
Therefore, the nontrivial fixed point is a saddle point.
  \item ${\bar{\xi}_2 > 0,\quad \left(\frac{72}{\bar{\xi}_2}\right)^{1/4}<|\bar{\varphi}_*|<\sqrt{\frac{9}{2\bar{\xi}_2}\left(1+\sqrt{1+\frac{56\bar{\xi}_2}{9}}\right)}}$\\
  In this case, both the eigenvalues are real and negative, $\lambda_{\pm}<0$. Therefore, the nontrivial fixed point is a stable fixed point (a sink).
  \item ${\bar{\xi}_2 > 0,\quad \bar{\varphi}_*>\sqrt{\frac{9}{2\bar{\xi}_2}\left(1+\sqrt{1+\frac{56\bar{\xi}_2}{9}}\right)}}$\\
In this case, the eigenvalues become complex, $\lambda_{\pm} = \alpha \pm i \beta$ with $\alpha<0$. Therefore, the nontrivial fixed point is a stable spiral fixed point (a spiral sink).
\end{enumerate}
The phase portraits for these four cases are shown in Fig.~\ref{fig:phaseportraits}.
Among these, of particular interest are the cases with a saddle point (1. and 2. above) as they may be considered as an alternative to inflation without the problem of graceful exit. The stable de Sitter cases (3. and 4.) may have some relevance, e.g., as dark energy, but we shall not discuss further these cases below.

\section{Ultra-slow-roll near the fixed point}\label{sec:USR}
Let us introduce the Hubble slow-roll parameters,
{\small
\begin{align}
\epsilon_1 \equiv -\frac{\mathring{\bar{H}}}{\bar{H}^2}
\,,\qquad
\epsilon_{i>1} \equiv \frac{\mathring{\epsilon}_{i-1}}{\bar{H}\epsilon_{i-1}}
\,.
\end{align}
}These are also written using the dimensionful variables as
$\epsilon_1=-\dot H/H^2$, $\epsilon_{i>1}=\dot\epsilon_{i-1}/H\epsilon_{i-1}$; see our definition of the variables \eqref{eqn:DimlessVars}.
It is also convenient to define the following parameters:
{\small
\begin{align}
\sigma_1 \equiv \bar{H}\mathring{\bar{\xi}}
=\bar{H}\bar{\xi}_{,\bar{\varphi}}\mathring{\bar{\varphi}}
\,,\qquad
\sigma_{i>1} \equiv \frac{\mathring{\sigma}_{i-1}}{\bar{H}\sigma_{i-1}}
\,,
\end{align}
}as well as the field acceleration parameter $f$,
{\small
\begin{align}
f \equiv -\frac{1}{3\bar{H}\mathring{\bar{\varphi}}}
\left(\frac{d\mathring{\bar{\varphi}}}{d\bar{t}}\right)
\,.
\end{align}
}Finally, we define $\delta$ through $f \equiv 1-\delta$.

Then, from the Klein-Gordon equation, we see that
{\small
\begin{align}
\delta = -\frac{-V_{,\varphi}+\frac{3}{2}H^2(\dot{H}+H^2)\xi_{,\varphi}}{3H\dot{\varphi}}
\,.
\end{align}
}Therefore, near the nontrivial fixed point, we naturally have the ultra-slow-roll condition, as $\delta\approx 0$ \cite{Inoue:2001zt,Kinney:2005vj}. See also Refs.~\cite{Pattison:2018bct,Dimopoulos:2017ged,Cheng:2018qof,Syu:2019uwx,Martin:2012pe,Mooij:2015yka,Firouzjahi:2018vet,Pattison:2019hef,Pattison:2021oen,Figueroa:2020jkf} for recent studies on ultra-slow-roll inflation.

In terms of the slow-roll parameters, the Klein-Gordon equation is given by
{\small
\begin{align}
3H\dot{\varphi}^2\delta + \dot{V}+\frac{3}{2}H^3M_{\rm P}^2(1-\epsilon_1)\sigma_1=0
\,,
\end{align}
}while, from the time derivative of the Friedmann equation, we obtain
{\small
\begin{align}
\frac{\dot{\varphi}^2}{H^2 M_{\rm P}^2}
=2\epsilon_1-\frac{1}{2}\sigma_1+\frac{1}{2}\sigma_1\sigma_2-\frac{1}{2}\sigma_1\epsilon_1\,.
\end{align}
}No approximation has been used.

From the above two equations, we get
{\small
\begin{align}
\delta = 1 - \frac{\epsilon_1}{3}
+\frac{4\epsilon_1\epsilon_2-\sigma_1\sigma_2+\sigma_1\sigma_2^2+\sigma_1\sigma_2\sigma_3-\sigma_1\sigma_2\epsilon_1-\sigma_1\epsilon_1\epsilon_2}
{6(4\epsilon_1-\sigma_1+\sigma_1\sigma_2-\sigma_1\epsilon_1)}
\,.
\end{align}
}Near the nontrivial fixed point, the inflaton velocity is small; i.e., $\sigma_1 \ll 1$. Furthermore, the condition for inflation is given by $\epsilon_1 \ll 1$.
So, imposing $\{\epsilon_1,\sigma_1\} \ll 1$, and the ultra-slow-roll condition, $\delta \approx 0$, we can express the second Hubble slow-roll parameter $\epsilon_2$ in terms of $\epsilon_1$ and $\sigma_i$ as follows:
{\small
\begin{align}
\epsilon_2 \approx -6 + \frac{\sigma_1}{4\epsilon_1}\big[
6-\sigma_2(5+\sigma_2+\sigma_3)
\big]\,.
\label{eqn:eps2ito}
\end{align}
}

Near the nontrivial fixed point, in terms of $x$ and $y$ defined via Eq.~\eqref{eqn:xANDy}, the first Hubble slow-roll parameter, $\epsilon_1$, is given by
{\small
\begin{align}
\epsilon_1 &= -\frac{\mathring{\bar{H}}}{\bar{H}^2}
\approx
\frac{(72-\bar{\xi}_2\bar{\varphi}_*^4)}{4\bar{\varphi}_*(6+\bar{\varphi}_*^2)}x
-\frac{4\sqrt{6}}{\bar{\varphi}_*|\bar{\varphi}_*|}y
\,,
\end{align}
}up to the first order in $\epsilon$.
In order to have inflation, we require $0 < \epsilon_1 < 1$, and thus,
{\small
\begin{align}
\frac{(72-\bar{\xi}_2\bar{\varphi}_*^4)|\bar{\varphi}_*|}{16\sqrt{6}(6+\bar{\varphi}_*^2)}x
\lesssim y \lesssim
\frac{\left[(72-\bar{\xi}_2\bar{\varphi}_*^4)x-4\bar{\varphi}_*(6+\bar{\varphi}_*^2)\right]|\bar{\varphi}_*|}{16\sqrt{6}(6+\bar{\varphi}_*^2)}
\,.
\end{align}
}Similarly, one can find the higher-order Hubble slow-roll parameters. 
The $\sigma_1$ parameter, up to the first order in $\epsilon$, becomes
{\small
\begin{align}
\sigma_1 &\approx
-\frac{4\sqrt{6}}{\bar{\varphi}_* |\bar{\varphi}_*|}y
\,.
\end{align}
}Indeed, $\epsilon_1$ and $\sigma_1$ are small parameters in the vicinity of the nontrivial fixed point. Furthermore, near the fixed point, we have $\delta \approx 0$. In terms of $x$ and $y$,
{\small
\begin{align}
\delta \approx
\left(\frac{|\bar{\varphi}_*|(72-\bar{\xi}_2\bar{\varphi}_*^4)}{12\sqrt{6}(6+\bar{\varphi}_*^2)}\right)\frac{x}{y}
\,,
\end{align}
}up to the leading order in $\epsilon$. Note that the leading order is the zeroth order. We focus on the region in the $x$--$y$ plane, where $\delta \approx 0$; we choose $\delta < 0.1$. Thus, the following condition should hold:
{\small
\begin{align}
y \gtrsim \left(
\frac{(72-\bar{\xi}_2\bar{\varphi}_*^4)|\bar{\varphi}_*|}{1.2\sqrt{6}(6+\bar{\varphi}_*^2)}
\right)x\,.
\end{align}
}In numerical analyses, we shall use the expression of $\delta$, expanded up to the first order in $\epsilon$.

We finish this section by finding the relation between $aH$ and the conformal time $\tau$.
We first note that
{\small
\begin{align}
\tau = \int \frac{da}{a^2H}
=-\frac{1}{aH}+\int\frac{\epsilon_1}{a^2H}da\,,
\end{align}
}where we used integration by parts.
Similarly, the second term becomes
{\small
\begin{align}
\int \frac{\epsilon_1}{a^2H}da &=
-\frac{\epsilon_1}{aH}+\int\frac{\epsilon_1\epsilon_2}{a^2H}da + \mathcal{O}\left(\epsilon_1^2\right)\,.
\end{align}
}Substituting Eq.~\eqref{eqn:eps2ito}, we obtain
{\small
\begin{align}
\int\frac{\epsilon_1\epsilon_2}{a^2H}da &\approx
-6\int\frac{\epsilon_1}{a^2H}da 
\nonumber\\
&\quad
+ \frac{3}{2}\int\frac{\sigma_1}{a^2H}\left(1-\frac{5}{6}\sigma_2-\frac{1}{6}\sigma_2^2-\frac{1}{6}\sigma_2\sigma_3\right)da\,.
\end{align}
}Collecting $\int (\epsilon_1/a^2H)da$, we get
{\small
\begin{align}
\int \frac{\epsilon_1}{a^2H}da &\approx 
-\frac{\epsilon_1}{7aH} 
\nonumber\\
&\quad
+ \frac{3}{14}\int\frac{\sigma_1}{a^2H}\left(1-\frac{5}{6}\sigma_2-\frac{1}{6}\sigma_2^2-\frac{1}{6}\sigma_2\sigma_3\right)da\,,
\end{align}
}and thus,
{\small
\begin{align}
\tau &\approx 
-\frac{1}{aH}
-\frac{\epsilon_1}{7aH}
\nonumber\\
&\quad
+\frac{3}{14}\int\frac{\sigma_1}{a^2H}\left(1-\frac{5}{6}\sigma_2-\frac{1}{6}\sigma_2^2-\frac{1}{6}\sigma_2\sigma_3\right)da\,.
\end{align}
}

Now, let us look at the $\sigma_1$ term,
{\small
\begin{align}
\int \frac{\sigma_1}{a^2H} &=
-\frac{\sigma_1}{aH} + \int \frac{\sigma_1\sigma_2}{a^2H}da + \mathcal{O}\left(\sigma_1\epsilon_1\right)\,.
\end{align}
}Thus,
{\small
\begin{align}
\tau &\approx 
-\frac{1}{aH} - \frac{\epsilon_1}{7aH} - \frac{3\sigma_1}{14aH} 
\nonumber\\
&\quad
+\frac{3}{84}\int\frac{\sigma_1\sigma_2}{a^2H}da
-\frac{3}{84}\int\frac{\sigma_1\sigma_2^2}{a^2H}da
-\frac{3}{84}\int\frac{\sigma_1\sigma_2\sigma_3}{a^2H}da\,.
\end{align}
}From the $\sigma_1\sigma_2$ term, we see that
{\small
\begin{align}
\int \frac{\sigma_1\sigma_2}{a^2H}da &=
-\frac{\sigma_1\sigma_2}{aH} 
\nonumber\\
&\quad
+\int\frac{1}{a^2H}\Big(
\sigma_1\sigma_2^2+\sigma_1\sigma_2\sigma_3
\Big)da+
\mathcal{O}\left(\epsilon_1\sigma_1\right)\,.
\end{align}
}

Therefore, we obtain
{\small
\begin{align}
\tau \approx
-\frac{1}{aH}
-\frac{\epsilon_1}{7aH}
-\frac{3\sigma_1}{14aH}
-\frac{3\sigma_1\sigma_2}{84aH}\,.
\end{align}
} In other words,
{\small
\begin{align}
(aH) \approx -\frac{1}{\tau}\left[
1+\frac{1}{7}\epsilon_1 + \frac{3}{14}\sigma_1\left(1+\frac{1}{6}\sigma_2\right)\right]\,.
\end{align}
}

\section{Density fluctuation and gravitational waves from de Sitter fixed point}\label{sec:pert}

\subsection{Density fluctuation}
The second-order perturbed action for curvature perturbation is given by 
{\small
\begin{align}
S^{(2)}_s = \frac{M_{\rm P}^2}{2} \int d\tau \, d^{3}x \, A_{\zeta}^2 \Big[
 (\zeta^\prime)^2 +C_{\zeta}^2 (\zeta\partial^2\zeta)
\Big]\,,
\label{eqn:2OSA}
\end{align}
}where
{\small
\begin{align}
A_{\zeta}^2 &\equiv 2 a^2 \left(
\frac{1-\sigma_1/2}{1-3\sigma_1/4}
\right)^2 
\nonumber\\
&\quad\times
\left[
\epsilon_1 - \frac{\sigma_1}{4} + \frac{\sigma_1\sigma_2}{4}
-\frac{\sigma_1\epsilon_1}{4}+\frac{3\sigma_1^2}{16(1-\sigma_1/2)}
\right]
\,,\\
C_{\zeta}^2 &\equiv
1-\frac{a^2 \sigma_1^2}{2A_{\zeta}^2(1-3\sigma_1/4)^2}\left(
\epsilon_1 + \frac{\sigma_1}{8} - \frac{5}{8}\sigma_1\epsilon_1
-\frac{\sigma_1\sigma_2}{8}
\right)
\,.
\end{align}
}For the derivation, see Appendix~\ref{apdx:CPTGB}.
In terms of a new variable, $u(\tau,\mathbf{x}) = M_{\rm P}A_\zeta \zeta(\tau,\mathbf{x})$, known as the Mukhanov-Sasaki variable, the action \eqref{eqn:2OSA} can be recast into
{\small
\begin{align}
S_s^{(2)} = \frac{1}{2}\int d\tau d^3x \left[
(u^\prime)^2 + C_\zeta^2 (u\partial^2 u) + \frac{A_\zeta^{\prime\prime}}{A_\zeta}u^2
\right]\,.
\end{align}
}Varying the action gives the following equation of motion for the variable $u$:
{\small
\begin{align}
u^{\prime\prime} - C_\zeta^2 \partial^2 u - \frac{A_\zeta^{\prime\prime}}{A_\zeta}u = 0\,.
\end{align}
}We perform the Fourier transformation,
{\small
\begin{align}
u(\tau,\mathbf{x}) = \int \frac{d^3k}{(2\pi)^3}e^{i\mathbf{k}\cdot\mathbf{x}}u(\tau,\mathbf{k})\,.
\end{align}
}The equation of motion then becomes
{\small
\begin{align}\label{eqn:MSeqn1}
u^{\prime\prime}(\tau,\mathbf{k}) + \left(
C_\zeta^2 k^2
-\frac{A_\zeta^{\prime\prime}}{A_\zeta}
\right) u(\tau,\mathbf{k}) = 0\,,
\end{align}
}with $k\equiv |\mathbf{k}|$.

We follow the standard quantization procedure and define the conjugate momentum $\pi_u \equiv \partial\mathcal{L}_s^{(2)}/\partial u^\prime = u^\prime$. The field $u$ and the conjugate momentum $\pi_u$ are then promoted to operators denoted with a hat, $\hat{u}$ and $\hat{\pi}_u$, with the equal-time canonical commutation relations, $[\hat{u}(\tau,\mathbf{x}),\hat{\pi}_u(\tau,\mathbf{y})]=i\delta(\mathbf{x}-\mathbf{y})$ and $[\hat{u}(\tau,\mathbf{x}),\hat{u}(\tau,\mathbf{y})]=[\hat{\pi}_u(\tau,\mathbf{x}),\hat{\pi}_u(\tau,\mathbf{y})]=0$. In the Fourier space, expanding the Fourier mode by using the creation and annihilation operators as $\hat{u}(\tau,\mathbf{k}) = a_{\mathbf{k}} u_k(\tau) + a_{-\mathbf{k}}^\dagger u_k^*(\tau)$, we see that the mode function $u_k(\tau)$ satisfies the same equation \eqref{eqn:MSeqn1} as $u(\tau,\mathbf{k})$.

For the initial conditions, it is natural to take the vacuum state $|0\rangle$ satisfying $a_{\mathbf{k}}|0\rangle = 0$. 
Acting the Hamiltonian on the vacuum state, we find that
{\small
\begin{align}
\hat{H}|0\rangle &= \frac{1}{2}\int\frac{d^3k}{(2\pi)^3}\left[
(u_k^\prime)^2 + \left(k^2 C_\zeta^2 - \frac{A_\zeta^{\prime\prime}}{A_\zeta}\right) u_k^2
\right]^* a_{-\mathbf{k}}^\dagger a_{\mathbf{k}}^\dagger |0\rangle 
\nonumber\\
&\quad\quad
+ E_0 |0\rangle\,,
\end{align}
}where $E_0$ is the ground-state energy. 
Demanding that the vacuum state $|0\rangle$ is an eigenstate of the Hamiltonian in the $\tau\to -\infty$ limit, the mode function $u_k$ needs to satisfy
{\small
\begin{align}
(u_k^\prime)^2 + \left( k^2 C_\zeta^2 - \frac{A_\zeta^{\prime\prime}}{A_\zeta} \right) u_k^2\simeq 0,
\quad
\tau\rightarrow-\infty.
\end{align}
}
Thus,
{\small
\begin{align}\label{eqn:ukprimepm}
u_k^\prime \simeq \pm i \left( k^2 C_\zeta^2 - \frac{A_\zeta^{\prime\prime}}{A_\zeta} \right)^{1/2} u_k,
\quad
\tau\to -\infty.
\end{align}
}For the normalization condition, we require the Wronskian $W[u_k,u_k^*] \equiv i(u_k^* u_k^\prime - u_k u_k^{*\prime})$ be unity, so that the creation and annihilation operators satisfy the canonical commutation relations $[a_\mathbf{k}, a_{\mathbf{k}^\prime}^\dagger] = (2\pi)^3 \delta(\mathbf{k} - \mathbf{k}^\prime)$. 
In the subhorizon limit, 
the $A''_\zeta/A_\zeta$ term is subdominant.
Then, Eq.~\eqref{eqn:ukprimepm} becomes
{\small
\begin{align}
u_k^\prime \approx \pm i C_\zeta k u_k\,.
\end{align}
}Now using the normalization condition $W[u_k,u_k^*] = 1$ and choosing the positive frequency mode, the mode function in the subhorizon limit is found to be
{\small
\begin{align}\label{eqn:WKBs}
\lim_{\tau\rightarrow -\infty} u_k(\tau) = \frac{1}{\sqrt{2C_\zeta k}}e^{-iC_\zeta k \tau}
\,,
\end{align}
}which is the standard WKB solution on the Bunch-Davies vacuum.

For the quadratic potential, we see that
{\small
\begin{align}
\sigma_2 \approx \frac{\sigma_1 - 4\epsilon_1}{\sigma_1}\,,
\end{align}
}up to the zeroth order in the small slow-roll parameters, $\epsilon_1$ and $\sigma_1$. Therefore, up to the first order,
{\small
\begin{align}
A_\zeta^2 \approx \frac{a^2}{2}\big(
4\epsilon_1 -\sigma_1 + \sigma_1\sigma_2
\big) \approx 0\,,
\end{align}
}and the leading-order contribution to $A_\zeta^2$ is the second order in $\epsilon_1$ and $\sigma_1$.
Under the ultra-slow-roll condition, the Mukhanov-Sasaki equation can be approximated as
{\small
\begin{align}
u_k^{\prime\prime} + \left(
C_\zeta^2 k^2
-\frac{\nu^2 - 1/4}{\tau^2}
\right) u_k = 0\,,
\label{eqn:MSeqnUSRscalar}
\end{align}
}where
{\small
\begin{align}
\nu &\approx
\sqrt{\frac{1}{4}+2\mathcal{A}^{(0)}}\bigg\{
1 + \frac{1}{1/4+2\mathcal{A}^{(0)}}
\nonumber\\
&\quad\quad\quad\quad\times
\left[
\mathcal{A}^{(1)}+\mathcal{A}^{(0)}\left(
\frac{2}{7}\epsilon_1 + \frac{3}{7}\sigma_1\left(
1+\frac{\sigma_2}{6}
\right)
\right)
\right]
\bigg\}\,,
\end{align}
}up to the first order in the small slow-roll parameters, $\epsilon_1$ and $\sigma_1$, and
{\small
\begin{align}
C_\zeta^2 \approx 1 + \frac{6\epsilon_1\sigma_1}{4\epsilon_1-3\sigma_1}\,.
\end{align}
}The expressions for $\mathcal{A}^{(0)}$ and $\mathcal{A}^{(1)}$ are summarized in Appendix~\ref{apdx:ExpressionsSummary}.
The solution of the Mukhanov-Sasaki equation \eqref{eqn:MSeqnUSRscalar} is then given by the Hankel function,
{\small
\begin{align}
u_k(\tau) = \frac{\sqrt{\pi}}{2}
e^{i\frac{\pi}{2}\left(\nu+\frac{1}{2}\right)}
\sqrt{-\tau}
H_{\nu}^{(1)}(-C_\zeta k \tau)\,,
\end{align}
}with the normalization fixed by the solution in the subhorizon limit \eqref{eqn:WKBs}.

The power spectrum,
{\small
\begin{align}
\mathcal{P}_\zeta =
\frac{k^3}{2\pi^2}\bigg\vert\frac{u_k}{M_{\rm P}A_\zeta}\bigg\vert^2
\,,
\end{align}
}is given, in the superhorizon limit, by
{\small
\begin{align}
\lim_{(C_\zeta k/aH)\rightarrow \infty}\mathcal{P}_{\zeta} &\approx
2^{2\nu-3}
\left(\frac{\Gamma(\nu)}{\Gamma(3/2)}\right)^2
\left(\frac{H}{2\pi}\right)^2
\left(\frac{C_\zeta k}{aH}\right)^{3-2\nu}
\nonumber\\
&\quad\times
\left[\frac{(1+\epsilon_1/7+3\sigma_1(1+\sigma_2/6)/14)^{1-2\nu}}
{C_\zeta^3 A_\zeta^2/a^2}
\right]
\,.
\label{eqn:curvaturePowSpec}
\end{align}
}The spectral index is given by
{\small
\begin{align}
n_s &= 1 + \frac{d\ln\mathcal{P}_\zeta}{d\ln k} = 4-2\nu\,.
\label{eqn:curvatureSpecIndex}
\end{align}
}Another scalar observable is the running of the spectral index, $\alpha_s$, which is defined as
{\small
\begin{align}
\alpha_s \equiv \frac{dn_s}{d\ln k}\,.
\label{eqn:curvatureSpecIndexRunning}
\end{align}
}

In terms of $x$ and $y$ defined via Eq.~\eqref{eqn:xANDy}, we see that the curvature power spectrum \eqref{eqn:curvaturePowSpec} up to the leading order is given by
{\small
\begin{align}
\mathcal{P}_\zeta \approx
2^{-3+2\nu}
\left(\frac{H}{2\pi}\right)^2
\left[\frac{\Gamma(\nu)}{\Gamma(3/2)}\right]^2
\frac{1}{\bar{A}_\zeta^{2}}
\left(\frac{k}{aH}\right)^{3-2\nu}
\,,
\end{align}
}where
{\small
\begin{align}
\bar{A}_\zeta^2 &\approx
\frac{6(6+\bar{\varphi}_*^2)}{\bar{\varphi}_*^4}y^2
\,,
\end{align}
}and
{\small
\begin{align}
\nu &\approx
\frac{1}{2}\sqrt{\frac{126+9\bar{\varphi}_*^2-\bar{\xi}_2\bar{\varphi}_*^4}{6+\bar{\varphi}_*^2}}
\,,
\end{align}
}is the leading-order contribution of the quantity $\nu$. At the horizon crossing, $k \approx aH$ (note that $C_\zeta \approx 1$ up to the leading order), we thus obtain
{\small
\begin{align}
\mathcal{P}_\zeta \approx
2^{-3+2\nu}
\left(\frac{H}{2\pi}\right)^2
\left[\frac{\Gamma(\nu)}{\Gamma(3/2)}\right]^2
\frac{1}{\bar{A}_\zeta^{2}}
\,.
\end{align}
}Let us compare the scalar power spectrum at a point, denoted by $*$, with the one at another point, denoted by $\odot$. We find
{\small
\begin{align}
\frac{\mathcal{P}_{\zeta}^\odot}{\mathcal{P}_{\zeta}^*}
\approx
\frac{\bar{A}_{\zeta}^{2*}}{\bar{A}_{\zeta}^{2\odot}}
\,.
\end{align}
}As we approach the nontrivial fixed point, $|y|$ decreases, and thus $\bar{A}_\zeta^2$ becomes smaller. In other words, depending on the initial condition of an inflationary trajectory, an enhancement of the curvature power spectrum may occur. A sufficient enhancement is known to lead to the formation of primordial black holes. We shall investigate this aspect in a separate paper \cite{Kawai:2021edk}.

The spectral index \eqref{eqn:curvatureSpecIndex} is given by
{\small
\begin{align}
n_s &=
4-\sqrt{\frac{126+9\bar{\varphi}_*^2-\bar{\xi}_2\bar{\varphi}_*^4}{6+\bar{\varphi}_*^2}}
\nonumber\\
&\quad
+\frac{1}{8\bar{\varphi}_* |\bar{\varphi}_*| (6+\bar{\varphi}_*^2)(126+9\bar{\varphi}_*^2-\bar{\xi}_2\bar{\varphi}_*^4)}
\sqrt{\frac{126+9\bar{\varphi}_*^2-\bar{\xi}_2\bar{\varphi}_*^4}{6+\bar{\varphi}_*^2}}
\nonumber\\
&\quad\times
\bigg[
|\bar{\varphi}_*|\big(
12096 + 288\bar{\varphi}_*^2 - 288\bar{\xi}_2\bar{\varphi}_*^4
+12\bar{\xi}_2\bar{\varphi}_*^6 + 3\bar{\xi}_2^2\bar{\varphi}_*^8
\nonumber\\
&\quad\quad
+24\bar{\xi}_3\bar{\varphi}_*^5 + 4\bar{\xi}_3\bar{\varphi}_*^7
\big)x
-72\sqrt{6}\left(72-\bar{\xi}_2\bar{\varphi}_*^4\right)y
\bigg]
\,.
\label{eqn:SSIxy}
\end{align}
}The running of the spectral index \eqref{eqn:curvatureSpecIndexRunning} is, up to the first order in $\epsilon$,
{\small
\begin{align}
\alpha_s &\approx
\frac{\sqrt{3/2}\sqrt{(126+9\bar{\varphi}_*^2-\bar{\xi}_2\bar{\varphi}_*^4)/(6+\bar{\varphi}_*^2)}}{4|\bar{\varphi}_*|(6+\bar{\varphi}_*^2)^2(-9\bar{\varphi}_*(14+\bar{\varphi}_*^2)+\bar{\xi}_2\bar{\varphi}_*^5)}
\nonumber\\
&\quad\times
\bigg[
3\sqrt{6}|\bar{\varphi}_*|\left(72-\bar{\xi}_2\bar{\varphi}_*^4\right)^2x
-(6+\bar{\varphi}_*^2)
\nonumber\\
&\quad\quad
\times\Big[
3\left(
96(96+\bar{\varphi}_*^2) - 4\bar{\xi}_2\bar{\varphi}_*^4(42-\bar{\varphi}_*^2)
+\bar{\xi}_2^2\bar{\varphi}_*^8
\right)
\nonumber\\
&\quad\quad\quad
+4\bar{\xi}_3\bar{\varphi}_*^5(6+\bar{\varphi}_*^2)
\Big]y
\bigg]
\,.
\label{eqn:SSIRxy}
\end{align}
}

\subsection{Gravitational waves}
The second-order perturbed action for tensor mode is given by
{\small
\begin{align}
S^{(2)}_t = \frac{M_{\rm P}^2}{8} \int d\tau \, d^3x \, A_t^2 \Big[
h^\prime_{ij}h^\prime_{ij} 
+ C_t^2 h_{ij} \partial^2 h_{ij}
\Big]\,,
\label{eqn:2OTA}
\end{align}
}where
{\small
\begin{align}
A_t^2 &\equiv
a^2\left(
1-\frac{\sigma_1}{2}
\right)
\,,\\
C_t^2 &\equiv
\frac{a^2}{A_t^2}\left(
1-\frac{\sigma_1}{2}(\sigma_2+\epsilon_1)
\right)
\,.
\end{align}
}For the derivation, see Appendix~\ref{apdx:CPTGB}.
Similarly to the scalar curvature perturbation, we introduce the Mukhanov-Sasaki variable $v^\pm$ by
{\small
\begin{align}
h_{ij} = \frac{\sqrt{2}}{A_t M_{\rm P}} \sum_{\pm}
v^\pm e_{ij}^\pm\,,
\end{align}
}where the summation is over the $\pm$ tensor polarization, and $\epsilon_{ij}^\pm$ are the polarization tensors. We follow the same procedure as before: we perform the Fourier transformation, quantize the theory, expand in terms of the mode function $v_k^\pm$, and impose the Bunch-Davies vacuum. The mode function $v_k^\pm$ then satisfies the following Mukhanov-Sasaki equation:
{\small
\begin{align}
(v_k^\pm)^{\prime\prime}
+\left(
C_t^2 k^2 - \frac{A_t^{\prime\prime}}{A_t}
\right)v_k^\pm = 0\,.
\end{align}
}From now on, we drop the superscript $\pm$ in the mode function.

Under the ultra-slow-roll condition, we get the following equation of motion for the tensor mode:
{\small
\begin{align}
v_k^{\prime\prime} + \left(
C_t^2 k^2 
- \frac{\nu_t^2-1/4}{\tau^2}
\right)v_k \approx 0\,,
\end{align}
}where
{\small
\begin{align}
C_t \approx
1+\frac{1}{4}(1-\sigma_2)\sigma_1
\,,
\end{align}
}and
{\small
\begin{align}
\nu_t \approx
\frac{3}{2}-\frac{1}{7}\epsilon_1
+\frac{1}{84}\sigma_1\left(
24-17\sigma_2-7\sigma_2^2-7\sigma_2\sigma_3
\right)
\,,
\end{align}
}up to the first order.
The solution is again given by the Hankel function,
{\small
\begin{align}
v_k(\tau) = \frac{\sqrt{\pi}}{2}e^{i\frac{\pi}{2}\left(\nu_t+\frac{1}{2}\right)}\sqrt{-\tau}
H_{\nu_t}^{(1)}(-C_t k \tau)\,,
\end{align}
}with the boundary conditions set by the Bunch-Davies vacuum.

The tensor power spectrum,
{\small
\begin{align}
\mathcal{P}_T = 
\frac{2k^3}{\pi^2}\bigg\vert\frac{v_k}{M_{\rm P}A_t}\bigg\vert^2\,,
\end{align}
}is given, in the superhorizon limit, by
{\small
\begin{align}
\lim_{(C_t k/aH)\rightarrow \infty}\mathcal{P}_{T} &\approx
2^{2\nu_t-1}
\left(\frac{\Gamma(\nu_t)}{\Gamma(3/2)}\right)^2
\left(\frac{H}{2\pi}\right)^2
\left(\frac{C_t k}{aH}\right)^{3-2\nu_t}
\nonumber\\
&\quad\times
\left[\frac{(1+\epsilon_1/7+3\sigma_1(1+\sigma_2/6)/14)^{1-2\nu_t}}
{C_t^3 A_t^2/a^2}
\right]
\,.
\end{align}
}

The tensor-to-scalar ratio $r$ is given by
{\small
\begin{align}
r &\equiv \frac{\mathcal{P}_T}{\mathcal{P}_\zeta}
\nonumber\\
&\approx
 \frac{2^{-\sqrt{1+8\mathcal{A}^{(0)}}}\pi}{\left[\Gamma\left(
\sqrt{1/4+2\mathcal{A}^{(0)}}
\right)\right]^2}
\left(
\frac{k}{aH}
\right)^{-3+\sqrt{1+8\mathcal{A}^{(0)}}}
\nonumber\\
&\quad\times
\sigma_1\left(
4\epsilon_1 + \sigma_1 + 2\sigma_1\sigma_2
\right)
\,.
\end{align}
}Note that the leading order is the second order in the small slow-roll parameters. This is due to the fact that $A_\zeta^2 = 0$ up to the first order.
In terms of $x$ and $y$, we obtain the following expression for the tensor-to-scalar ratio:
{\small
\begin{align}
r \approx
\frac{3\pi\left(
6+\bar{\varphi}_*^2
\right)2^{4-\sqrt{\frac{126+9\bar{\varphi}_*^2-\bar{\xi}_2\bar{\varphi}_*^4}{6+\bar{\varphi}_*^2}}}}
{\bar{\varphi}_*^4\left[\Gamma\left(\frac{1}{2}\sqrt{\frac{126+9\bar{\varphi}_*^2-\bar{\xi}_2\bar{\varphi}_*^4}{6+\bar{\varphi}_*^2}}\right)\right]^2}
y^2
\,.
\label{eqn:TTSratioxy}
\end{align}
}

\section{Observational Constraints}\label{sec:constraints} 

\begin{figure*}[ht!]
\centering
\includegraphics[scale=0.8]{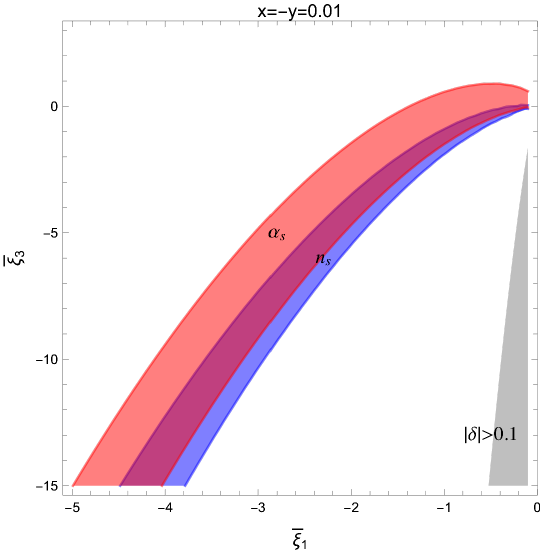}\qquad
\includegraphics[scale=0.8]{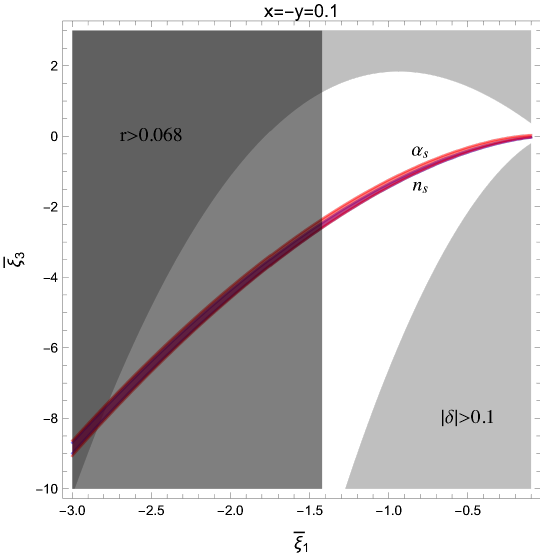}
\caption{Allowed model parameter space. The second model parameter $\bar{\xi}_2$ is fixed by the center value of the spectral index and the value of $\bar{\xi}_1$. The initial condition is chosen to be $x=-y=0.01$ (left) and $x=-y=0.1$ (right). The blue (red) region is the Planck-allowed bound for the (running) spectral index. The light gray region is excluded by the ultra-slow-roll condition, $|\delta|<0.1$. The dark gray region is excluded by the Planck bound on the tensor-to-scalar ratio, $r<0.068$.}
\label{fig:resultxy}
\end{figure*}

Having obtained the expressions for inflationary observables, such as the spectral index \eqref{eqn:SSIxy}, the running of the spectral index \eqref{eqn:SSIRxy}, and the tensor-to-scalar ratio \eqref{eqn:TTSratioxy}, let us now use the latest Planck data \cite{Akrami:2018odb} to constrain the model,
{\small
\begin{align}
n_s &= 0.9658 \pm 0.0040 
\,,\nonumber\\
\alpha_s &= -0.0066\pm 0.0070 
\,,\label{eqn:Planck2018data}\\
r &< 0.068 
\,.\nonumber
\end{align}
}

The model parameters are $\bar{\xi}_i$ ($i=1,2,\cdots$) and $m$. In addition, there are two parameters, $x$ and $y$, for the initial conditions. Fixing the initial conditions in general requires a full picture of the scalar potential as well as the coupling function, which are expected to be given by an UV-complete theory. Our approach is phenomenological, and we are interested in possible phenomenological interpretations of the de Sitter fixed points of the cosmological solutions. Therefore, we will first fix $x$ and $y$ and see how the model parameters are constrained. We will then discuss the initial condition dependence.

\begin{figure*}[ht!]
\centering
\includegraphics[scale=0.8]{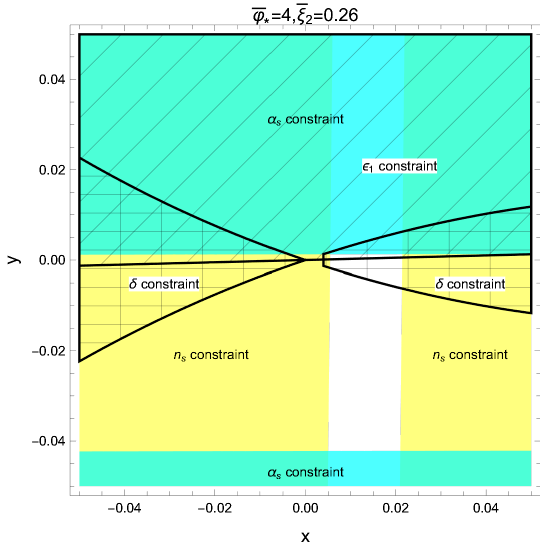}\qquad
\includegraphics[scale=0.8]{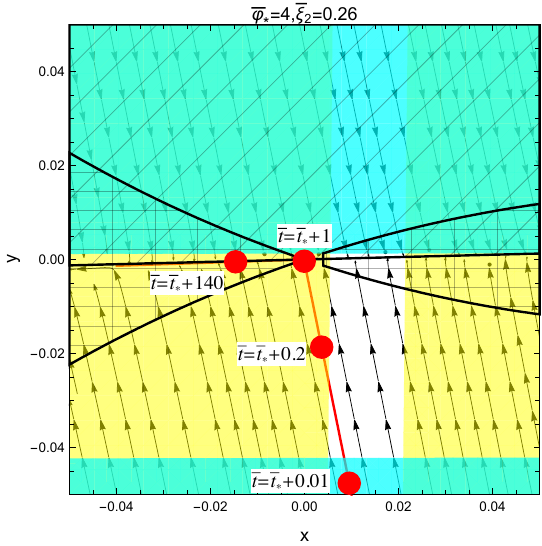}
\caption{The figure on the left panel shows the allowed region in the $x$--$y$ plane after imposing the Planck data \eqref{eqn:Planck2018data} for $\{\bar{\varphi}_*,\bar{\xi}_2\}=\{4,0.26\}$. The yellow region is excluded by the $n_s$ bounds, cyan region by $\alpha_s$, the boxed region by $\delta > 0.1$, and the meshed region by $\epsilon_1>1$.
The figure on the right panel shows the phase portrait and a particular trajectory. Here, $\bar{t}_*$ is a reference (dimensionless) time. We see that the inflaton spends much longer time in the vicinity of the nontrivial fixed point.}
\label{fig:result4}
\end{figure*}
\begin{figure*}[ht!]
\centering
\includegraphics[scale=0.6]{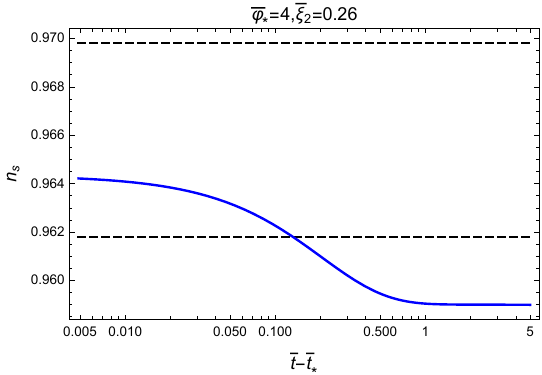}\;\;
\includegraphics[scale=0.6]{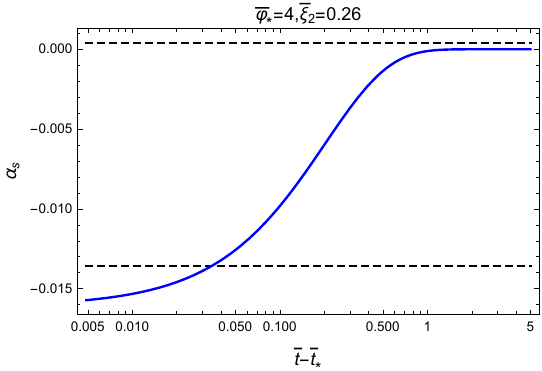}\;\;
\includegraphics[scale=0.6]{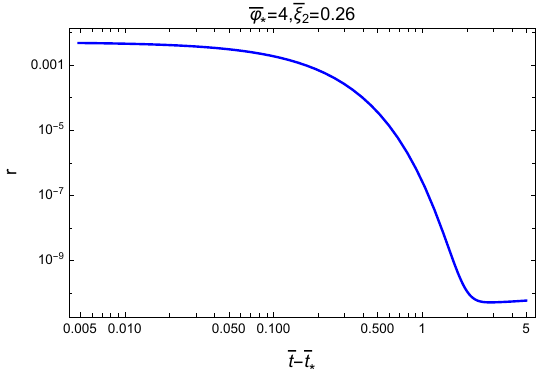}
\caption{Time evolutions of $n_s$, $\alpha_s$, and $r$, respectively, from left to right, for $\{\bar{\varphi}_*,\bar{\xi}_2\}=\{4,0.26\}$. Dashed lines indicate the Planck bounds \eqref{eqn:Planck2018data}. While the spectral index $n_s$ decreases as the trajectory approaches the nontrivial fixed point, the running of the spectral index increases. Therefore, improvements of their bounds will help to understand better the characteristics of the nontrivial fixed point as well as the dynamics of the model. The extreme suppression of the tensor-to-scalar ratio is expected as $r\sim \mathcal{O}(\epsilon^2)$. A detection of large tensor-to-scalar ratio may rule out the model.}
\label{fig:evolution4}
\end{figure*}

The mass of the inflaton $m$ is determined in such a way that the normalization of the scalar power spectrum amplitude matches the Planck result \cite{Akrami:2018odb}, $A_s \approx 2.1\times 10^{-9}$ at the pivot scale $k=0.05\;{\rm Mpc}^{-1}$. The scale $k$ is related to the $e$-folding number between the horizon exit of the modes and the reheating. As we are only interested in the physics near the nontrivial fixed point, we consider the chosen potential $V$ and coupling function $\xi$ to be reliable only near the fixed point, and assume that the subsequent evolution of the inflaton dynamics justifies the comparison of the spectra of the previous Sec. \ref{sec:pert} and the Planck result. For the coefficients of the Taylor-expanded coupling function $\xi(\varphi)$, we ignore negligible terms that are higher than the third order; i.e., $\bar{\xi}_{i>3}=0$. We are thus left with three free parameters, $\bar{\xi}_1$, $\bar{\xi}_2$, and $\bar{\xi}_3$.

We first note that there exists a correlation between the value of the nontrivial fixed point $\bar{\varphi}_*$ and $\bar{\xi}_2$, from the zeroth-order correction in the spectral index \eqref{eqn:SSIxy}. One may express $\bar{\xi}_2$ in terms of $\bar{\varphi}_*$ and $n_s$ by using the zeroth-order expression of the spectral index,
{\small
\begin{align}
\bar{\xi}_2 &\approx
\frac{30 - 7\bar{\varphi}_*^2}{\bar{\varphi}_*^4}
+\frac{48+8\bar{\varphi}_*^2}{\bar{\varphi}_*^4}n_s
-\frac{6+\bar{\varphi}_*^2}{\bar{\varphi}_*^4}n_s^2\,.
\end{align}
}Recall that the nontrivial fixed point $\bar{\varphi}_*$ is given by $\bar{\xi}_1$ as shown in Eq.~\eqref{eqn:FPxi1}. We thus see that $\bar{\xi}_2$ is not an independent free parameter; it is a function of $n_s$ and $\bar{\xi}_1$. We use the center value of the spectral index to find $\bar{\xi}_2$ for a given $\bar{\xi}_1$.
Using the inflation condition, $0<\epsilon_1<1$, and the ultra-slow-roll condition, $\delta\approx0$, together with the Planck data \eqref{eqn:Planck2018data}, we constrain the remaining two free parameters, $\bar{\xi}_1$ and $\bar{\xi}_3$.

In Fig.~\ref{fig:resultxy}, the allowed region for the model parameters after imposing the five constraints is shown. Two sets of initial condition are considered: $x=-y=0.01$ (left panel of Fig.~\ref{fig:resultxy}) and $x=-y=0.1$ (right panel of Fig.~\ref{fig:resultxy}). We see that, for the case of $x=-y=0.01$, the model parameters are constrained mostly by the spectral index (blue) and the running of the spectral index (red). On the other hand, for the case of $x=-y=0.1$, the ultra-slow-roll condition, $|\delta|<0.1$, and the tensor-to-scalar ratio bound, $r<0.068$, kick in to exclude most of the parameter space, leaving a narrow allowed region. Improvements of the bound on the spectral index and its running will reduce the number of free parameters down to one, as $\bar{\xi}_3$ will be given by a function of $\bar{\xi}_1$. On the other hand, improvements of the bound on the tensor-to-scalar ratio will tightly constrain the model parameter $\bar{\xi}_1$.

Let us discuss the initial condition dependence. We fix the model parameters to see the dependence on $x$ and $y$.
Figure \ref{fig:result4} shows the allowed region in the $x$--$y$ plane as well as a particular trajectory for $\{\bar{\varphi}_*,\bar{\xi}_2\}=\{4,0.26\}$. In Fig.~\ref{fig:evolution4}, evolutions of $n_s$, $\alpha_s$, and $r$, following the particular trajectory, are shown. The Planck bounds \eqref{eqn:Planck2018data}, together with the inflation condition and the ultra-slow-roll condition, favors a small region of the initial condition space. From a particular trajectory shown in the right panel of Fig. \ref{fig:result4}, we see that the inflaton spends much longer time in the vicinity of the nontrivial fixed point.

As shown in Fig. \ref{fig:evolution4}, the spectral index $n_s$ decreases as the trajectory approaches the nontrivial fixed point, while the running of the spectral index increases. Improvements of their bounds will help to understand better the characteristics of the nontrivial fixed point as well as the dynamics of the model. The tensor-to-scalar ratio is extremely suppressed since $r\sim \mathcal{O}(\epsilon^2)$. Therefore, a detection of large tensor-to-scalar ratio may rule out the model.

\section{Final remarks}\label{sec:concl} 

We investigated phenomenological implications of nontrivial de Sitter fixed points in the presence of a nonminimal coupling of a massive scalar field to the Euler density by studying its dynamical system. We studied the behavior of cosmological solutions near the nontrivial de Sitter fixed points, which resembles the ultra-slow-roll inflation. A generic nonminimal coupling function is considered, and we showed the existence of observationally consistent parameter region.

Our model exhibits extremely small values of $r$. Future cosmic microwave background B-mode observations \cite{Remazeilles:2015hpa}, such as LiteBIRD \cite{Hazumi:2019lys,Matsumura:2013aja} and PIXIE \cite{Kogut:2011xw}, target to achieve the uncertainty on the tensor-to-scalar ratio $\Delta r\lesssim 10^{-3}$. Thus, finding of a large $r$ would disfavor our scenario.

Another interesting feature of our model is the enhancement of the curvature power spectrum. Such an enhancement may lead to primordial black hole formations. The primordial black holes may contribute a significant amount to the current dark matter relic density. Furthermore, the enhanced curvature power spectrum may source the tensor perturbation, giving rise to an interesting gravitational waves signal.

Finally, we comment on the swampland conjectures \footnote{
The swampland conjectures for the Gauss-Bonnet cosmology were discussed in Refs. \cite{Yi:2018dhl,Odintsov:2020zkl} (only the slow roll case was studied).
}. The de Sitter conjecture \cite{Obied:2018sgi} (see also Refs. \cite{Garg:2018reu,Ooguri:2018wrx} for a refined version), which is one of the swampland conjectures, states that $M_{\rm P}|V_{,\varphi}|/V \gtrsim c \sim \mathcal{O}(1)$. It immediately puts standard slow-roll single field inflation in danger \cite{Agrawal:2018own,Kinney:2018nny} as the tensor-to-scalar ratio, given by $r\approx 16\epsilon_V \gtrsim 8c^2$, where the potential slow-roll parameter is defined as $\epsilon_V \equiv (M_{\rm P}^2/2)(V_{,\varphi}/V)^2\gtrsim c^2/2$, is already excluded by the Planck data. The value of $c$ could be reduced down to $\mathcal{O}(0.1)$. The bound on the tensor-to-scalar ratio, however, is still too large, $r \gtrsim 0.08$.
In our setup, due to the presence of the coupling function $\xi(\varphi)$, near the nontrivial fixed point, we have the ultra-slow-roll regime. In this regime, the Hubble slow-roll parameter $\epsilon_1$ is related to the potential slow-roll parameter $\epsilon_V$ as $\epsilon_1 \approx \Xi\epsilon_V$, where $\Xi \equiv [12M_{\rm P}^2/(\xi_{,\varphi}V_{,\varphi}^2)][V_{,\varphi}+V^2\xi_{,\varphi}/(6M_{\rm P}^4)]$. Near the nontrivial fixed point, $|\Xi| \ll 1$, and thus, we have the desired inflation condition $\epsilon_1 < 1$ without violating the de Sitter conjecture.

\begin{acknowledgments}
This work was supported in part by the National Research Foundation of Korea Grant-in-Aid for Scientific Research Grant No.
NRF-2018R1D1A1B07051127, the NRF-JSPS Collaboration ``String Axion Cosmology" (S.K).
\end{acknowledgments} 

\appendix

\section{Cosmological perturbation theory in Gauss-Bonnet gravity}
\label{apdx:CPTGB}
Cosmological perturbation theory, including the Gauss-Bonnet coupling, was developed in Refs. \cite{Kawai:1997mf,Kawai:1999pw,Kawai:1998ab}. In this appendix, we derive second-order perturbed actions, \eqref{eqn:2OSA} and \eqref{eqn:2OTA}, for scalar and tensor perturbations.
We work in the comoving (uniform field) gauge and consider the ADM metric perturbation,
{\small
\begin{align}
ds^2 = -N^2 dt^2 + \gamma_{ij}(N^i dt + dx^i)(N^j dt + dx^j)\,,
\end{align}
}where $N$ and $N^i$ are the lapse and shift functions, and
{\small
\begin{align}
\gamma_{ij} = a^2e^{2\zeta}\left(
e^h
\right)_{ij}
\end{align}
}is the three spatial-dimensional metric.
Here, $\zeta$ is the curvature perturbation and $h_{ij}$ is the transverse-traceless tensor perturbation.

\subsection{Scalar perturbation}
Let us focus on the scalar perturbation first. Expanding the action \eqref{eqn:action} up to the second order with
{\small
\begin{align}
N = 1+\alpha \,,\quad
N^i = \partial^i \beta\,,\quad
\gamma_{ij} = a^2e^{2\zeta}\delta_{ij}\,,
\end{align}
}we obtain
{\small
\begin{align}
S^{(2)}_s = \int dt \, d^3x \, \mathcal{L}^{(2)}_s\,,
\end{align}
}where
{\small
\begin{align}
\mathcal{L}^{(2)}_s &=
a^3 \alpha^2 \left(
-\dot{H} M_{\rm P}^{2} +\frac{\ddot{\xi} H^{2}}{4}
+\frac{11}{4}\dot{\xi} H^{3} - 3H^{2} M_{\rm P}^{2}
+\frac{H \dot{H} \dot{\xi}}{2}
\right)
\nonumber\\
&\quad
+a^3 \alpha \dot{\zeta} \left(
6H M_{\rm P}^{2}  - \frac{9}{2}\dot{\xi} H^{2}
\right)
+a^3 \dot{\zeta}^2 \left(
-3 M_{\rm P}^{2} + \frac{3}{2}H \dot{\xi}
\right)
\nonumber\\
&\quad
+a \alpha \partial^{2}\zeta  \left(
-2 M_{\rm P}^{2} +H \dot{\xi}
\right)
+a \dot{\zeta} \partial^{2} \beta \left(
2 M_{\rm P}^{2} -H \dot{\xi}
\right)
\nonumber\\
&\quad
+a \zeta \partial^{2} \zeta \left(
\frac{\ddot{\xi}}{2} - M_{\rm P}^{2}
\right)
+a \alpha \partial^{2} \beta \left(
-2H M_{\rm P}^{2} +\frac{3}{2}\dot{\xi} H^{2}
\right)
\,.
\end{align}
}

Varying the action with respect to $\alpha$ and $\beta$ gives the following constraint equations:
{\small
\begin{align}
0 &=
2a^2 \alpha \left(
-\dot{H} M_{\rm P}^{2} +\frac{\ddot{\xi} H^{2}}{4}
+\frac{11}{4}\dot{\xi} H^{3} - 3H^{2} M_{\rm P}^{2}
+\frac{H \dot{H} \dot{\xi}}{2} 
\right)
\nonumber\\
&\quad
+a^2 \dot{\zeta} \left(
6H M_{\rm P}^{2}  - \frac{9}{2}\dot{\xi} H^{2}
\right)
+\partial^{2}\zeta \left(
-2 M_{\rm P}^{2} +H \dot{\xi}
\right)
\nonumber\\
&\quad
+\partial^{2}\beta\left(
-2H M_{\rm P}^{2} +\frac{3}{2}\dot{\xi} H^{2}
\right)
\,,\\
0 &=
\dot{\zeta} \left(
2 M_{\rm P}^{2} -H \dot{\xi}
\right)
+\alpha\left(
-2H M_{\rm P}^{2} +\frac{3}{2}\dot{\xi} H^{2}
\right)
\,,
\end{align}
}which, in terms of the slow-roll parameters, are rewritten as
{\small
\begin{align}
0 &= a^2 H^2 \alpha \left(
-3 + \epsilon_1 + \frac{11}{4}\sigma_1
+\frac{\sigma_1\sigma_2}{4} 
-\frac{\sigma_1\epsilon_1}{4}
\right)
\nonumber\\
&\quad
+3a^2 H\dot{\zeta} \left(
1 - \frac{3}{4}\sigma_1
\right)
-H \partial^2 \beta \left(
1-\frac{3}{4}\sigma_1
\right)
\nonumber\\
&\quad
-\partial^2 \zeta \left(
1 - \frac{1}{2}\sigma_1
\right)
\,,\label{eqn:ADMbeta}\\
0 &= \dot{\zeta}\left(
1 - \frac{1}{2}\sigma_1
\right)
-H \alpha \left(
1 - \frac{3}{4}\sigma_1
\right)
\,.\label{eqn:ADMalpha}
\end{align}
}From Eq.~\eqref{eqn:ADMalpha}, we see that
{\small
\begin{align}
\alpha = \left(
\frac{1 - \sigma_1/2}{1-3\sigma_1/4}
\right)\frac{\dot{\zeta}}{H}
\,.
\end{align}
}We substitute the equation for $\alpha$ into Eq.~\eqref{eqn:ADMbeta} to obtain
{\small
\begin{align}
\frac{\partial^{2} \beta}{a^2} &=
\frac{1-\sigma_1/2}{(1-3\sigma_1/4)^2}
\bigg(
\epsilon_1 - \frac{7}{4}\sigma_1 + \frac{27}{16}\sigma_1^2
+\frac{1}{4}\sigma_1\sigma_2 - \frac{1}{4}\sigma_1\epsilon_1
\bigg)\dot{\zeta}
\nonumber\\
&\quad
-\left(
\frac{1-\sigma_1/2}{1-3\sigma_1/4}
\right)\frac{\partial^{2} \zeta}{a^2 H}
\,.
\end{align}
}

Therefore, the second-order scalar Lagrangian is given by
{\small
\begin{align}
\mathcal{L}^{(2)}_s &=
a^3 M_{\rm P}^2 \left(
\frac{1-\sigma_1/2}{1-3\sigma_1/4}
\right)^2 \bigg[
-3+\epsilon_1+\frac{11}{4}\sigma_1
+\frac{\sigma_1\sigma_2}{4}-\frac{\sigma_1\epsilon_1}{4}
\nonumber\\
&\quad
+3\frac{(1-3\sigma_1/4)^2}{1-\sigma_1/2}
\bigg]\dot{\zeta}^2
-2aM_{\rm P}^2\frac{(1-\sigma_1/2)^2}{1-3\sigma_1/4}\frac{\dot{\zeta}}{H}\partial^2 \zeta
\nonumber\\
&\quad
-aM_{\rm P}^2\left(1-\frac{1}{2}\sigma_1\sigma_2-\frac{1}{2}\sigma_1\epsilon_1\right)\zeta\partial^2\zeta
\,.
\end{align}
}Using $\dot{\zeta}\partial^{2}\zeta (\cdots) = -(1/2)\zeta\partial^2 \zeta \partial_{t}(\cdots)$ up to the total derivative, we find that
{\small
\begin{align}
\mathcal{L}^{(2)}_s &=
a^3 M_{\rm P}^2 \left(
\frac{1-\sigma_1/2}{1-3\sigma_1/4}
\right)^2 \bigg[
\epsilon_1 - \frac{\sigma_1}{4} + \frac{\sigma_1\sigma_2}{4}
-\frac{\sigma_1\epsilon_1}{4}
\nonumber\\
&\quad
+\frac{3\sigma_1^2}{16(1-\sigma_1/2)}
\bigg]\dot{\zeta}^2
+aM_{\rm P}^2\left(
\frac{1-\sigma_1/2}{1-3\sigma_1/4}
\right)^2
\bigg[
\epsilon_1 - \frac{\sigma_1}{4} 
\nonumber\\
&\quad
+ \frac{\sigma_1\sigma_2}{4}
-\frac{\sigma_1\epsilon_1}{4}+\frac{3\sigma_1^2}{16(1-\sigma_1/2)}
\nonumber\\
&\quad
-\frac{\sigma_1^2}{4(1-\sigma_1/2)^2}\left(
\epsilon_1 + \frac{\sigma_1}{8} - \frac{5}{8}\sigma_1\epsilon_1
-\frac{\sigma_1\sigma_2}{8}
\right)
\bigg]
\zeta\partial^2\zeta
\,.
\end{align}
}

Let us define
{\small
\begin{align}
A_{\zeta}^2 &\equiv 2 a^2 \left(
\frac{1-\sigma_1/2}{1-3\sigma_1/4}
\right)^2 
\nonumber\\
&\quad\times
\left[
\epsilon_1 - \frac{\sigma_1}{4} + \frac{\sigma_1\sigma_2}{4}
-\frac{\sigma_1\epsilon_1}{4}+\frac{3\sigma_1^2}{16(1-\sigma_1/2)}
\right]
\,,\\
C_{\zeta}^2 &\equiv
1-\frac{a^2 \sigma_1^2}{2A_{\zeta}^2(1-3\sigma_1/4)^2}\left(
\epsilon_1 + \frac{\sigma_1}{8} - \frac{5}{8}\sigma_1\epsilon_1
-\frac{\sigma_1\sigma_2}{8}
\right)
\,.
\end{align}
}Then, in terms of the conformal time, the second-order action is given by
{\small
\begin{align}
S^{(2)}_s = \frac{M_{\rm P}^2}{2} \int d\tau \, d^{3}x \, A_{\zeta}^2 \Big[
 (\zeta^\prime)^2 +C_{\zeta}^2 (\zeta\partial^2\zeta)
\Big]\,.
\label{eqn:2OActionScalar}
\end{align}
}

\subsection{Tensor perturbation}
For the tensor perturbation, we expand the action \eqref{eqn:action} up to the second order with
{\small
\begin{align}
\gamma_{ij} = a^2\left(
\delta_{ij} + h_{ij} + \frac{1}{2}h_{ik}h_{kj}
\right)
\,.
\end{align}
}We then obtain
{\small
\begin{align}
S^{(2)}_t = \int dt \, d^3x \, \mathcal{L}^{(2)}_t\,,
\end{align}
}with
{\small
\begin{align}
\mathcal{L}^{(2)}_t &=
\frac{a^3M_{\rm P}^2}{8}\bigg[
\left(1 - \frac{\dot{\xi}H}{2M_{\rm P}^2}\right)
\dot{h}_{ij}\dot{h}_{ij}
+\left(1 - \frac{\ddot{\xi}}{2M_{\rm P}^2}\right)
\frac{h_{ij} \partial^2 h_{ij}}{a^2}
\bigg]
\,,
\end{align}
}where we used the transverse-traceless condition, $\partial_j h_{ij} = h_{ii}=0$.
Note that $\dot{\xi}H/M_{\rm P}^2 = \sigma_1$ and $\ddot{\xi}/M_{\rm P}^2 = \sigma_1(\sigma_2+\epsilon_1)$.
Let us define
{\small
\begin{align}
A_t^2 \equiv
a^2\left(
1-\frac{\sigma_1}{2}
\right)
\,,\quad
C_t^2 \equiv
\frac{a^2}{A_t^2}\left(
1-\frac{\sigma_1}{2}(\sigma_2+\epsilon_1)
\right)
\,.
\end{align}
}Then, in terms of the conformal time, the second-order action is given by
{\small
\begin{align}
S^{(2)}_t = \frac{M_{\rm P}^2}{8} \int d\tau \, d^3x \, A_t^2 \Big[
h^\prime_{ij}h^\prime_{ij} 
+ C_t^2 h_{ij} \partial^2 h_{ij}
\Big]\,.
\label{eqn:2OActionTensor}
\end{align}
}

\section{Expressions of $\mathcal{A}^{(0)}$ and $\mathcal{A}^{(1)}$}
\label{apdx:ExpressionsSummary}
In this appendix, we summarize expressions of $\mathcal{A}^{(0)}$ and $\mathcal{A}^{(1)}$,
{\small
\begin{align}
\mathcal{A}^{(0)} &=
\frac{1}{8(4\epsilon_1-(1-\sigma_2)\sigma_1)^2}\Bigg\{
16\left(8+\epsilon_2(6+\epsilon_2+2\epsilon_3)\right)\epsilon_1^2
\nonumber\\
&\quad
+\bigg[
8+\sigma_2^4 + 4(1+\sigma_3)\sigma_2^3-(3-\sigma_3^2+2\sigma_3\sigma_4)\sigma_2^2
\nonumber\\
&\quad
-2\sigma_2\left(5+\sigma_3(2+\sigma_3+\sigma_4)\right)
\bigg]\sigma_1^2
\nonumber\\
&\quad
-8\bigg[
8+\epsilon_2^2(1-\sigma_2)+\epsilon_2(1-\sigma_2)(3+\epsilon_3-\sigma_2)
+\epsilon_2\sigma_2\sigma_3
\nonumber\\
&\quad
-\sigma_2\left(
5+\sigma_2^2+\sigma_2(2+3\sigma_3)+\sigma_3(2+\sigma_3+\sigma_4)
\right)
\bigg]\epsilon_1\sigma_1
\Bigg\}
\,,\\
\mathcal{A}^{(1)} &=
\frac{1}{16(4\epsilon_1-(1-\sigma_2)\sigma_1)^3}\Bigg\{
-256(2+\epsilon_2)\epsilon_1^4
\nonumber\\
&\quad
-64\left(-2(3+\epsilon_2)+(2+\epsilon_2)\sigma_2\right)\epsilon_1^3\sigma_1
\nonumber\\
&\quad
+\sigma_2\bigg[
3+4\sigma_2^4 + 10\sigma_3 + 4\sigma_2^3\sigma_3 - \sigma_2^2(9+13\sigma_3)
\nonumber\\
&\quad
+3\sigma_3(\sigma_3+\sigma_4)-\sigma_2(-2+\sigma_3+3\sigma_3\sigma_4)
\bigg]\sigma_1^4
\nonumber\\
&\quad
-16\bigg[
6+\epsilon_2\left(7+\epsilon_3(2+\sigma_2)+\sigma_2(7+\sigma_2+\sigma_3)\right)
\nonumber\\
&\quad
-\sigma_2\left(
13+6\sigma_2^2+\sigma_2(17+8\sigma_3)+\sigma_3(6+\sigma_3+\sigma_4)
\right)
\bigg]\epsilon_1^2\sigma_1^2
\nonumber\\
&\quad
-4\bigg[
-2+\epsilon_2^2(-2+\sigma_2+\sigma_2^2) -\epsilon_2\sigma_2\sigma_3(5+\sigma_2)
\nonumber\\
&\quad
-\epsilon_2(1-\sigma_2)\left(
6+(4-\sigma_2)\sigma_2+\epsilon_3(2+\sigma_2)
\right)
\nonumber\\
&\quad
+\sigma_2\big(
14-9\sigma_2^3-2\sigma_2^2(7+5\sigma_3)
+4\sigma_3(4+\sigma_3+\sigma_4)
\nonumber\\
&\quad
+\sigma_2(11+15\sigma_3-\sigma_3\sigma_4)
\big)
\bigg]\epsilon_1\sigma_1^3
\Bigg\}\,.
\end{align}
}


\end{document}